\documentclass[aps,pre,twocolumn,superscriptaddress,longbibliography,nobibnotes,nodoi,nourl,noeprint]{revtex4-2}
\usepackage{amsmath,amssymb,physics,bm,siunitx}
\usepackage{xcolor,graphicx,soul}
\usepackage[colorlinks=true,citecolor=black,urlcolor=black,linkcolor=black]{hyperref}

\def\rms{\rm\scriptscriptstyle}
\DeclareMathOperator*{\intpart}{\rm int}

\begin{document}
	
\title{Single-file transport of binary hard-sphere mixtures through periodic potentials} 

\author{David Vor{\'a}{\v c}}
\affiliation{Charles University, Faculty of Mathematics and Physics, Department of Macromolecular Physics, V Hole\v{s}ovi\v{c}k\'{a}ch 2, 
CZ-18000 Praha 8, Czech Republic}

\author{Philipp Maass} 
\email{maass@uos.de}
\affiliation{Universit\"{a}t Osnabr\"{u}ck, Faculty of Mathematics, Informatics and Physics, Institute of Physics, 
Barbarastra{\ss}e 7, D-49076 Osnabr\"uck, Germany}

\author{Artem Ryabov}
\email{rjabov.a@gmail.com}
\affiliation{Charles University, Faculty of Mathematics and Physics, Department of Macromolecular Physics, V Hole\v{s}ovi\v{c}k\'{a}ch 2, 
CZ-18000 Praha 8, Czech Republic}

\date{August 9, 2023} 

\begin{abstract}
Single-file transport occurs in various scientific fields, including diffusion through nanopores, nanofluidic devices, and cellular processes. We here investigate the impact of polydispersity on particle currents for single-file Brownian motion of hard spheres, when they are driven through periodic potentials by a constant drag force. Through theoretical analysis and extensive Brownian dynamics simulations, we unveil the behavior of particle currents for random binary mixtures. The particle currents show a recurring pattern in dependence of the hard-sphere diameters and mixing ratio. We explain this recurrent behavior by showing that a basic unit cell exists in the space of the two hard-sphere diameters. Once the behavior of an observable inside the unit cell is determined, it can be inferred for any diameter. The overall variation of particle currents with the mixing ratio and hard-sphere diameters is reflected by their variation in the limit where the system is fully covered by hard spheres. In this limit, the currents can be predicted analytically. Our analysis explains the occurrence of pronounced maxima and minima of the currents by changes of an effective potential barrier for the center-of-mass motion.
\end{abstract} 
\maketitle

\section{Introduction}
Single-file dynamics refers to the collective motion in a many-particle system,
where the particles cannot overtake each other because of their interactions and spatial confinements. Examples of such process are diffusion in pores of zeolites \cite{Hahn/etal:1996, VanDeVoorde/Sels:2017, Chmelik/etal:2023}, colloidal particles confined to move
to move in one-dimensional circular channels \cite{Wei/etal:2000},
particle motion through carbon nanotubes \cite{Zeng/etal:2018}, in nanofluidic devices \cite{Ma/etal:2015} and in mesoporous materials \cite{Hartmann:2005, Humphrey/etal:2001}.
Single-file transport often occurs in cell biology, e.g.\ in the movement of motor proteins \cite{Kolomeisky:2013, Kolomeisky:2015}, 
in protein synthesis by ribosomes
\cite{MacDonald/etal:1968, Erdmann-Pham:2020}, 
and in ion motion through narrow channels \cite{Hille:2001}.  

Recently, the Brownian asymmetric simple exclusion process (BASEP) was introduced as a paradigmatic model 
of single-file diffusion in a periodic potential \cite{Lips/etal:2018, Lips/etal:2019, Antonov/etal:2022a}.
A striking feature of the BASEP is the sensitive dependence of its
physics on the particle diameter $\sigma$.  This is present even when
ignoring a dependence of the particle mobility $\mu$ on $\sigma$. Such
a dependence naturally occurs in fluid environments, where it can be
accounted for by Stokes' friction law, $\mu\propto 1/\sigma$, for a
single particle. More complicated forms may be relevant in
the single-file dynamics of interacting particles due to hydrodynamic interactions. 

Here we study the impact of polydispersity on the transport behavior in the BASEP.
Polydispersity can hardly be fully avoided in experiments and it can be highly relevant in single-file transport, as, for example, in
biological membranes and channels, in which transport of various types of proteins, ions, or
other biomolecules takes place. Such membranes and channels are highly selective, with only
a few types of particles being able to pass through.
Underlying selection mechanisms often involve particle size.

We focus on a binary mixture of particles with hard-core
interactions, i.e.\ two types of hard spheres with diameters $\sigma_{\rm A}$
and $\sigma_{\rm B}>\sigma_{\rm A}$. Our primary interest is to understand how
particle currents vary with the mixing ratio of the two types of particles and the hard-sphere
diameters. Under single-file conditions, the particle currents also depend 
on the specific ordering of particles. We will show, however, that they exhibit a 
typical behavior for a given mixing ratio, and will therefore consider currents 
averaged over
orderings. 

We furthermore tackle commensurability effects, where the hard-sphere diameters 
are integer multiples of the period length $\lambda$ of the
external potential. Due to these effects, a basic unit cell can be defined
in the $\sigma_{\rm A}\sigma_{\rm B}$-plane encompassing all hard-sphere diameters. By studying observables within this unit cell, one can predict their behavior anywhere else in the plane using transformation laws.

The transformation laws imply that it is interesting to study single-file transport of hard-sphere
diameters under single-file conditions, where the diameters would actually allow the particle to overtake each other, i.e.\
where sums of two hard-sphere diameters are smaller than the pore diameter.
By transformation to larger hard-sphere diameters, single file constraints become fulfilled.

The paper is organized as follows. After presenting the BASEP mixture model in Sec.~\ref{sec:model}, 
we derive in Sec.~\ref{sec:mapping}
a mapping between probabilities of Brownian paths of many-body systems with particles of different
hard-sphere diameters and discuss its implication on single-file constraints. 
The transformation between currents resulting from this mapping is provided in 
Sec.~\ref{sec:mapping-of-currents}, which serves as a useful basis for explaining repeating features in the dependence of 
particle currents on particle size. Results from extensive Brownian dynamics simulations
for the particle currents are presented in Sec.~\ref{sec:simulated-currents}. We interpret their dependence on the mixing ratio and particle
diameters by considering the limit of a full coverage, where the currents can be calculated based on the Brownian motion 
of the center of mass in an effective potential.

\section{BASEP mixture model}
\label{sec:model}
\begin{figure}[t!]	
\centering
\includegraphics[width=1\columnwidth]{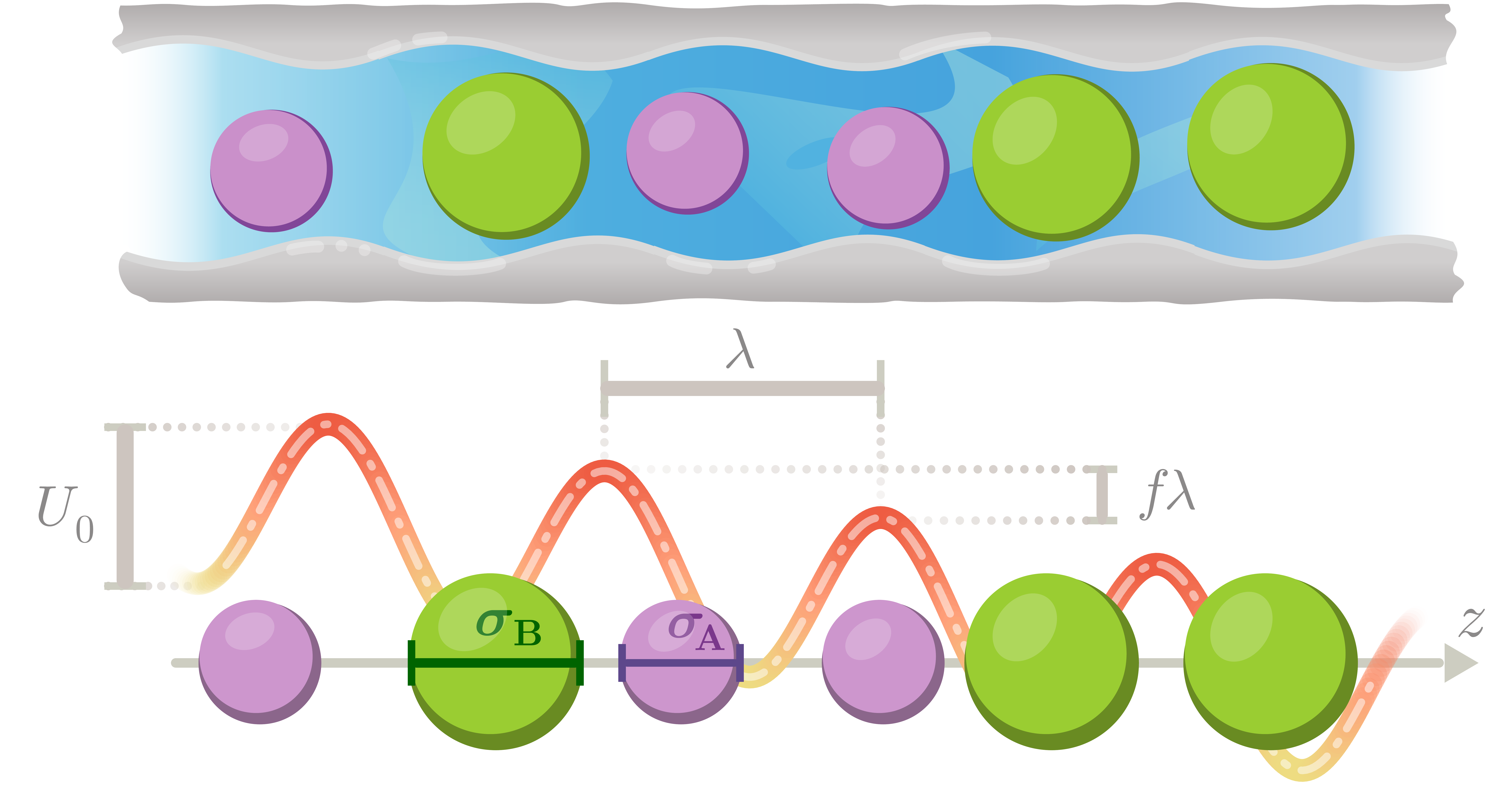}
\caption{Schematic of a corrugated channel with a mixture of particles of two sizes (top). In the model, hard spheres of two diameters $\sigma_{\rm A}$ and $\sigma_{\rm B}$ are driven along a line
by a constant drag force $f$  through a periodic potential $U(z)$  with wavelength $\lambda$ (bottom).}
\label{fig:model-illustration}
\end{figure}

We consider a binary mixture of hard spheres A and B with particle diameters $\sigma_{\rm A}$ and 
$\sigma_{\rm B}\ge\sigma_{\rm A}$, which are driven through a narrow 
pore with a periodic structure in an aqueous environment as illustrated in Fig.~\ref{fig:model-illustration}. 
The overdamped Brownian motion of a sphere $i$ of 
type $\alpha$, $\alpha={\rm A}$ or B, is described by the Langevin equation
\begin{equation}
\frac{\dd z_i}{\dd t} = \mu_\alpha \left( f_\alpha - \frac{\dd U_\alpha}{\dd z_i} \right) + \sqrt{2D_\alpha}\,\xi_i(t)\,,
\label{eq:langevin}
\end{equation} 
where $z_i(t)$ is the position of the sphere,
$f_\alpha>0$ is a constant drag force, 
and $U_\alpha(z)$ a periodic external potential reflecting the interactions with the pore walls;
$\mu_\alpha$ and $D_\alpha$ are the bare mobility and diffusion coefficient of particles of type $\alpha$, where
$D_\alpha=k_{\rm B} T\mu_\alpha$. In accord with the Stokes law for the drag force, we set
\begin{equation}
D_\alpha=\frac{\lambda}{\sigma_\alpha}D
\label{eq:Dratio}
\end{equation}
for the diffusion coefficients of the two particle types, 
where $D$ is a bare diffusion coefficient of a reference sphere with diameter $\lambda$. 
The total number of particles is $N$, and
the $\xi_i(t)$, $i=1,\ldots, N$, are Gaussian white noise processes satisfying $\langle \xi_i(t) \rangle =0$, $\langle \xi_i(t)\xi_j(t') \rangle =\delta_{ij}\delta(t-t')$. 

For simplicity, we assume  in this study the drag forces $f_\alpha$ and the potentials $U_\alpha(z)$ to be equal for both types of particles, 
$f_{\rm A}=f_{\rm B}=f$ and $U_{\rm A}(z)=U_{\rm B}(z)=U(z)$, where
\begin{equation}
U(z) = \frac{U_0}{2} \cos\left(\frac{2\pi z}{\lambda} \right)\,.
\label{eq:external-potential}
\end{equation}
Here, $\lambda$ is the wavelength of the potential and $U_0$ the barrier between potential wells.

The hard-sphere interaction implies the constraints 
\begin{equation}
|z_{i+1}-z_i| \geq \frac{\sigma_{i+1}+\sigma_i}{2}\,.
\label{eq:hs-constraints}
\end{equation}

For analyzing fundamental diagrams of particle currents as a function of particle densities, 
we investigate the particle motion in a closed system with $M$ wells under periodic boundary conditions. Accordingly, the system 
length is
\begin{equation}
L=M\lambda\,.
\end{equation}
Note that if the particle labels are assigned such as \mbox{$z_1<\ldots<z_N$} at some time instant, the distance between the last and first particle
must be smaller than $L-(\sigma_1+\sigma_N)/2$, i.e.\  $(z_N-z_1)<L-(\sigma_1+\sigma_N)/2$.
The number density of particles is
\begin{equation}
\rho=\frac{N}{L}=\frac{N_{\rm A}+N_{\rm B}}{L}=\rho_{\rm A}+\rho_{\rm B}\,,
\end{equation}
where $N$ is the total number of particles, and $N_\alpha$ and $\rho_\alpha=N_\alpha/L$
are the numbers and number densities of particles of type $\alpha$.
The mixing ratio of the two particle types is 
\begin{equation}
x=\frac{N_{\rm B}}{N}\,.
\end{equation}
Accordingly, $\rho_{\rm A}=(1-x)\rho$ and  $\rho_{\rm B}=x\rho$. 
Since the coverage $N_{\rm A}\sigma_{\rm A}+N_{\rm B}\sigma_{\rm B}$ 
by the particles must be smaller than the system length $L$,
we must have $\rho_{\rm A}\sigma_{\rm A}+\rho_{\rm B}\sigma_{\rm B}=\rho(1-x)\sigma_{\rm A}-\rho x\sigma_{\rm B}<1$, i.e.\
the range of $x$ is limited to 
\begin{equation}
x<\frac{1/\rho-\sigma_{\rm A}}{\sigma_{\rm B}-\sigma_{\rm A}}\,.
\label{eq:x-constraint}
\end{equation}

In the following, we adopt 
$k_{\rms B}T$, $\lambda$, and $\lambda^2/D$ as units of energy, length, and time, respectively. The potential barrier is set to
$U_0=6$, i.e.\ it is six times larger than the thermal energy. 

As the drag force $f$ increases, the potential barriers gradually diminish, eventually vanishing at the critical force $f_c=\pi U_0/\lambda$,
where the local maxima of the tilted potential $[U(z)-fz]$ turn into saddle points. At $f=f_c$, $[U(z)-fz]$ has
vanishing first and second order derivatives, which yields $f_c=\pi U_0/\lambda$.
This critical force marks the transition from overcritical to undercritical tilting of the periodic potential.
We are interested here in the undercritical regime $f<f_c$ and have set $f=k_{\rm B}T/\lambda=f_c/6\pi$.

We simulated the time evolution according to Eq.~\eqref{eq:langevin} and subject to the constraints \eqref{eq:hs-constraints}
by using Scala's event-driven algorithm \cite{Scala/etal:2007, Scala:2012, Scala:2013} 
with a time-step $\Delta t = 0.4\times10^{-3}$. Particle currents $J_\alpha$ were sampled after the system has evolved into a nonequilibrium steady state. 
All simulations were carried out
for system sizes corresponding to $M=50$ potential wells. 

\section{Single-file constraints and mapping between path probabilities }
\label{sec:mapping}
The restriction of particle motion to one dimension in the model is motivated by considering the
motion to take place in a narrow pore under single-file condition, where particles cannot overtake each other. If the pore diameter is $d$, single-file motion for both the A and B spheres would occur for $2\sigma_{\rm A}>d$, as we defined 
$\sigma_{\rm B}>\sigma_{\rm A}$. Because $\sigma_{\rm B}$ must be smaller than $d$, this implies the single-file condition
\begin{equation}
\sigma_{\rm B}<2\sigma_{\rm A}\,.
\label{eq:single-file-condition}
\end{equation}
Studying situations by the model, where this condition is violated, seems to be unphysical.
However, as we will show now, it is still valuable to consider particle sizes that violate the single-file condition. This is because the current (and other quantities) in such systems can be mapped onto the current in a system with larger particle sizes that do satisfy the single-file conditions. 

Intuitively, such a mapping is imaginable due to the fact that a shifting of particle positions by a multiple integer of $\lambda$ does not change the external forces  in the periodic potential. Hence, one should obtain an equivalent system when increasing the distances between particles' centers of mass by integer multiples of 
$\lambda$. By increasing particle sizes, the interactions between particles can be kept the same, i.e.\ the new particle coordinates and sizes 
should obey the hard-sphere constraints~\eqref{eq:hs-constraints}. In addition, the change of positions and sizes should be accompanied by an appropriate change of the system length to accommodate all particles.

The mapping is strictly true 
when neglecting the difference between the diffusion coefficients of the 
two particle types as given by Eq.~\eqref{eq:Dratio}, and approximately when including this dependence, as we will 
see in Sec.~\ref{sec:simulated-currents}.

For $D_\alpha=D$,
the probability $P \big[\{z_i(t)\} ; L, \sigma_{\rm A},\sigma_{\rm B} \big]$ of finding a path (trajectory)
$\{z_i(t)\}$ in a system $\mathcal{S}$ with particle sizes $\sigma_\alpha$ is equal to the probability 
$P \big[\{z_i'(t)\} ; L', \sigma_{\rm A}',\sigma_{\rm B}' \big]$ of finding a corresponding path $\{z_i'(t)\}$
in a system $\mathcal{S}'$ with smaller particle sizes $\sigma_\alpha'$,
\begin{equation}
P \big[\{z_i(t)\} ; L, \sigma_{\rm A},\sigma_{\rm B} \big]=P \big[\{z_i'(t)\} ; L', \sigma_{\rm A}',\sigma_{\rm B}' \big]\,.
\label{eq:path-probabilites}
\end{equation}
This holds under the following conditions:
\begin{list}{}{\setlength{\leftmargin}{1.2em}\setlength{\rightmargin}{0em}
\setlength{\itemsep}{0ex}\setlength{\topsep}{0ex}}

\item[(i)] The path $\{z_i'(t)\}$ follows from the path $\{z_i(t)\}$ by a transformation specified below, see Eqs.~\eqref{eq:zi-transform1}
and \eqref{eq:zi-transform2}.

\item[(ii)] The particle diameters differ by a multiple integer of the wavelength $\lambda$, 
\begin{subequations}
\label{eq:sigma-transform}
\begin{align}
\sigma_{\rm A}'&=\sigma_{\rm A}-m_{\rm A}\lambda\,,\hspace{1em}m_{\rm A}\in\{1,2,\ldots,\intpart(\sigma_{\rm A}/\lambda)\}\,,
\label{eq:sigmaA-transform}\\
\sigma_{\rm B}'&=\sigma_{\rm B}-m_{\rm B}\lambda\,,\hspace{1em}m_{\rm B}\in\{1,2,\ldots,m_{\rm B}^{\rm max}\}\,,
\label{eq:sigmaB-transform}
\end{align}
\end{subequations}
where $m_{\rm B}^{\rm max}=\intpart(\sigma_{\rm B}/\lambda)$ if $\intpart(\sigma_{\rm B}/\lambda)$ is even, and
$m_{\rm B}^{\rm max}=\intpart(\sigma_{\rm B}/\lambda)-1$ otherwise.

\item[(iii)] 
The particle numbers in both systems $\mathcal{S}$ and $\mathcal{S}'$ are the same, but the system lengths transform as
\begin{equation}
L'=L-(m_{\rm A}N_{\rm A}+m_{\rm B}N_{\rm B})\lambda\,. 
\label{eq:L-transform}
\end{equation}
This corresponds to a mapping
\begin{align}
\rho'&=\frac{N}{L'}
=\frac{N}{L-(m_{\rm A}N_A+m_{\rm B}N_{\rm B})\lambda}
\nonumber\\&
=\frac{\rho}{1-[m_{\rm A}(1-x)+m_{\rm B}x]\rho\lambda}
\label{eq:rhoprime-rho}
\end{align}
between particle densities.

\end{list}

To show this, we proceed in two steps. First, we consider the case $m_{\rm A}=m_{\rm B}=m$. Second, we consider different $m_{\rm A}$ and $m_{\rm B}$ that are both even. By combining the results, we arrive at the conclusion in Eq.~\eqref{eq:path-probabilites}.

\subsection{Case $m_{\rm A}=m_{\rm B}$}
For $m_{\rm A}=m_{\rm B}=m$, we apply the coordinate transformation
\begin{equation}
z_i' = z_i-im\lambda
\label{eq:zi-transform1}
\end{equation}
to the Langevin equations
\eqref{eq:langevin}. 
The transformation leaves the
external forces $f^{\rms ext}(z_i)=-\dd U(z_i)/\dd z_i$ unaltered, 
$f^{\rms ext}(z_i') = f^{\rms ext}(z_i-im\lambda) = f^{\rms ext}(z_i)$, and 
the hard-sphere constraints~\eqref{eq:hs-constraints} become
$(z_{i+1}- z_i)=z'_{i+1}-z_i'+m\lambda\ge (\sigma_i+\sigma_{i+1})/2$. Hence, if we set
\begin{equation}
\sigma_\alpha'=\sigma_\alpha-m\lambda\,,
\end{equation} 
we obtain $(z'_{i+1}-z_i')\ge (\sigma_i'+\sigma_{i+1}')/2$,
i.e.\ analogous constraints. 

Moreover, the condition $(z_N-z_1) <[L-(\sigma_1+\sigma_N)/2]$ transforms into 
$(z_N'-z_1')<L' -Nm\lambda -(\sigma_1'+\sigma_N')/2$, corresponding to a change 
$L'\to L=L'-Nm\lambda$ of the system length. Mathematically, the corresponding change of length
ensures that the probability measure for the joint probability density of particle positions does not change.

Thus, Eq.~\eqref{eq:path-probabilites} is valid for $m_{\rm A}=m_{\rm B}$, if the particle diameters and
system length are transformed according to Eqs.~\eqref{eq:sigma-transform} and \eqref{eq:L-transform}.

\subsection{Case $m_{\rm A}\ne m_{\rm B}$}
For $m_{\rm A}\ne m_{\rm B}$, we start by considering both $m_{\rm A}$ and $m_{\rm B}$ to be even.
In that case, we apply the coordinate transformation
\vspace{-1ex}\begin{align*}
z_1'&= z_1-\frac{m_1}{2}\lambda\,,\\
z_2'&= z_2-\left(m_1+\frac{m_2}{2}\right)\lambda\,,\\[-1ex]
& \hspace{0.6em}\vdots
\end{align*}
\vspace{-3ex} i.e.\ \vspace*{1ex} 
\begin{equation}
z_j'=z_j-\left(\sum_{i=1}^{j-1}m_i+\frac{m_j}{2}\right)\lambda,\hspace{1em}j=1,\ldots, N\,.
\label{eq:zi-transform2}
\end{equation}
Because all $m_i$ are even, this transformation again leaves the
external forces $f^{\rms ext}(z_i)=-\dd U(z_i)/\dd z_i$ unaltered.
As for the  hard-sphere constraints~\eqref{eq:hs-constraints},
we obtain
$(z_{i+1}- z_i)=z'_{i+1}-z_i'-(m_{i+1}+m_i)\lambda/2\ge (\sigma_i+\sigma_{i+1})/2$.
Hence, if we set
$\sigma_\alpha'=\sigma_\alpha-m_\alpha\lambda$\,,
we again obtain analogous constraints $(z'_{i+1}-z_i')\ge (\sigma_i'+\sigma_{i+1}')/2$.

Using Eq.~\eqref{eq:zi-transform2}, the condition $(z_N-z_1) <[L-(\sigma_1+\sigma_N)/2]$
 transforms into $(z_N - z_1) < [L' -(\sigma_1'+\sigma'_N)/2-\sum_{j=1}^Nm_j\lambda
 =[L' -(\sigma_1'+\sigma'_N)/2-(N_{\rm A}m_{\rm A}+N_{\rm B}m_{\rm B})\lambda$, corresponding to a change 
$L\to L'=L-(N_{\rm A}m_{\rm A}+N_{\rm B}m_{\rm B})\lambda$ of the system length, in agreement with
Eq.~\eqref{eq:L-transform}.

\begin{figure}[t!]	
\centering
\includegraphics[width=0.7\columnwidth]{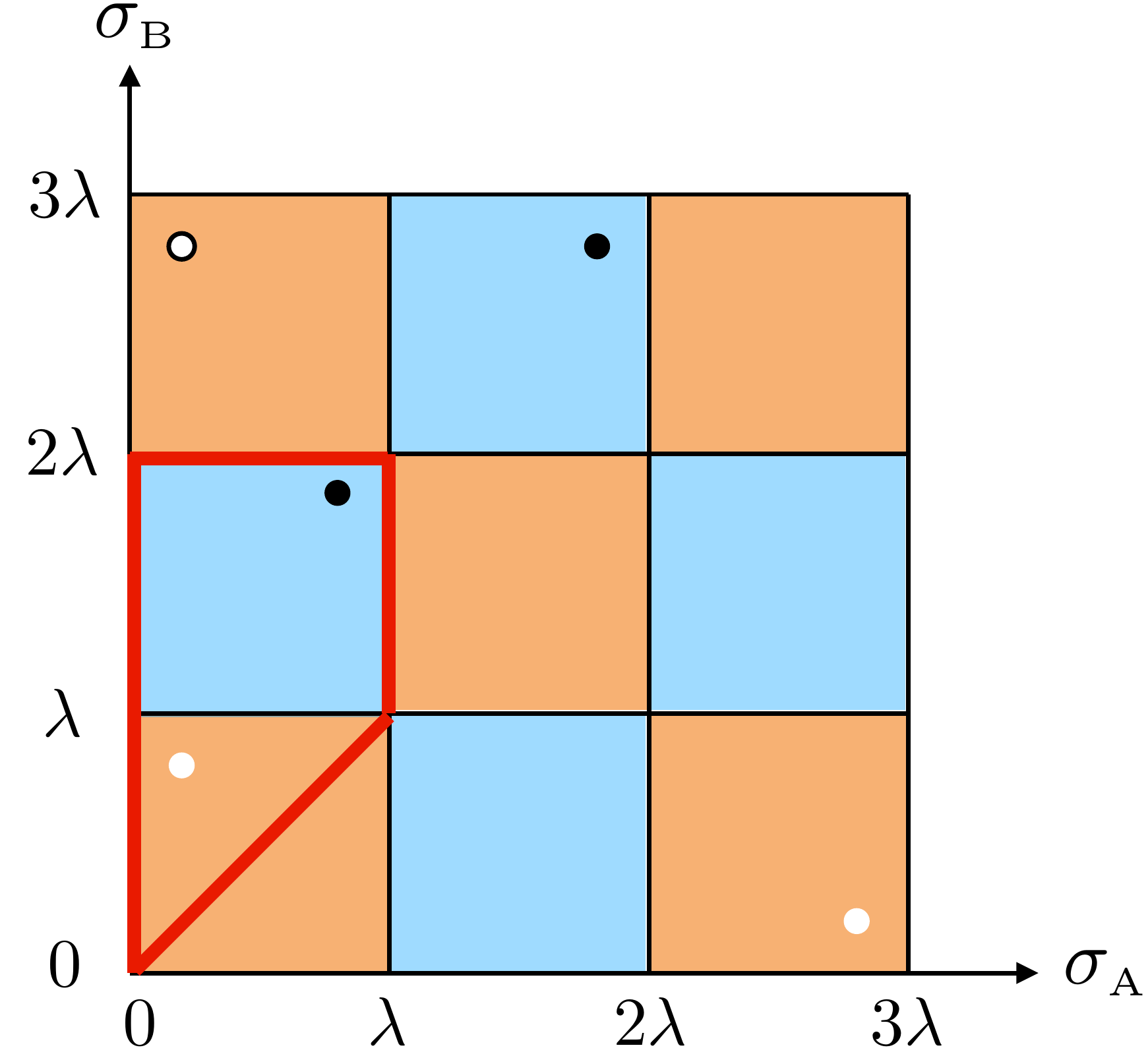}
\caption{Corresponding regions in the $\sigma_{\rm A}\sigma_{\rm B}$-plane: The value of an observable at any point in the blue (orange) areas of the $\sigma_{\rm A}\sigma_{\rm B}$-plane can be inferred from the knowledge
of the observable at a corresponding point in the blue (orange) area in the basic unit cell $\mathcal{C}$ 
bordered by the bold red line [see Eq.~\eqref{eq:unitcell}]. In the mapping between observables, the mixing ratio $x$
is kept fixed, the density transforms according to Eq.~\eqref{eq:rhoprime-rho}, and the particle diameters according to 
Eqs.~\eqref{eq:sigmaA-transform}, \eqref{eq:sigmaB-transform} for $\sigma_{\rm B}>\sigma_{\rm A}$.
Black circles illustrate a mapping for a point with $\sigma_{\rm B}>\sigma_{\rm A}$
onto a point in $\mathcal{C}$. A point with $\sigma_{\rm B}<\sigma_{\rm A}$ is first reflected at the 
diagonal $\sigma_{\rm B}=\sigma_{\rm A}$, and is then mapped onto a point in $\mathcal{C}$ by
applying Eqs.~\eqref{eq:sigmaA-transform}, \eqref{eq:sigmaB-transform}. This is illustrated by the white circles, where the framed white circle indicates the mirror image.} 
\label{fig:illustration-mapping}
\end{figure}	

\begin{figure*}[t!]	
\centering
\includegraphics[width=0.8\textwidth]{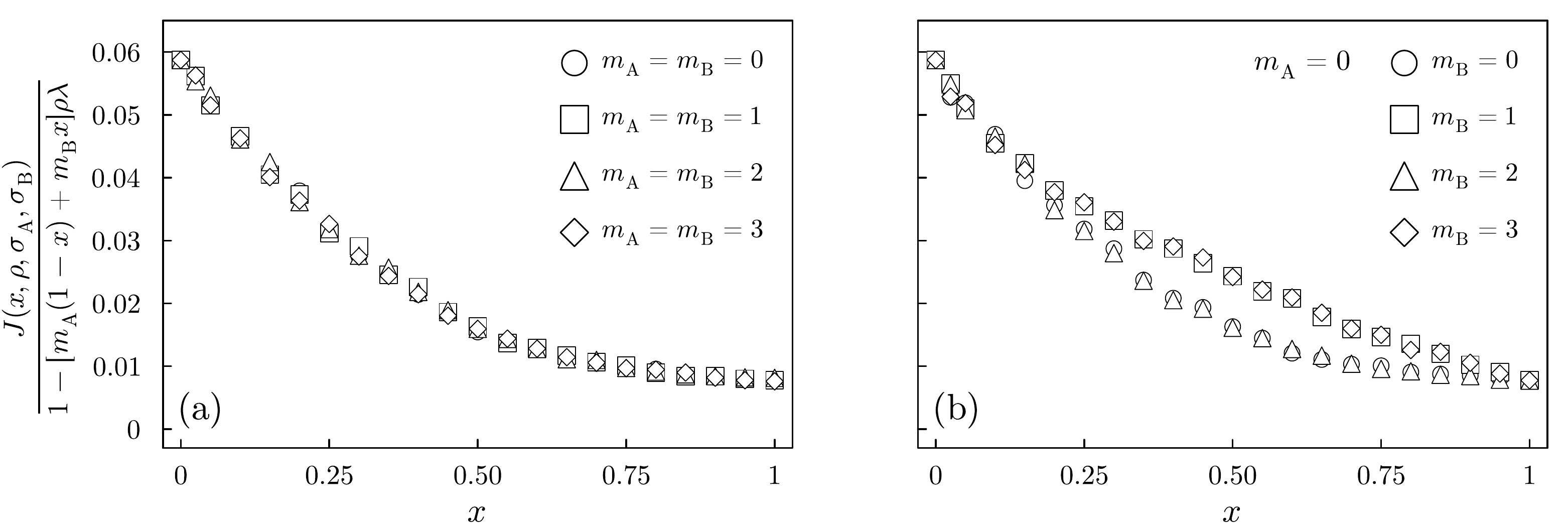}
\caption{Demonstration of the current mapping according to Eq.~\eqref{eq:mapping-J-2} for a fixed particle number $N=40$ and $\rho'=0.8$, implying $L'=N/\rho'=50$.
The particle diameters in the basic unit cells are $\sigma'_{\rm A}=0.4\lambda$ and $\sigma'_{\rm B}=0.7\lambda$.
In (a) the scaled currents fall onto a common master curve for different $m_{\rm A}=m_{\rm B}$. In (b) the mapping is shown
for $m_{\rm A}=0$ and different $m_{\rm B}$, where the currents fall onto two distinct master curves for even and odd $m_{\rm B}$.
The system length was varied with the $m_\alpha$ according to
$L=L'(1+[m_{\rm A}(1-x)+m_{\rm B}x]\rho'\lambda)$, giving $\rho$ corresponding to $\rho'$. For each mixing ratio $x$, the currents were averaged over 40 random sequences of the A and B particles.} 
\label{fig:current-maping}
\end{figure*}	

Let us now consider a system with some diameters $\sigma_{\rm A}<\sigma_{\rm B}$. Applying the transformation for
equal $m_\alpha=m$ with $m=\intpart(\sigma_{\rm A}/\lambda)$ then yields an ``equivalent system'' $\mathcal{S}'$ with
$\sigma_{\rm A}'\in[0,\lambda[$ and $\sigma_{\rm B}'=\sigma_{\rm B}-m\lambda$. If $m_{\rm B}'=\intpart(\sigma'_{\rm B}/\lambda)$
is even, a subsequent transformation with even $m_{\rm A}'=0$ and $m_B'$ gives an unaltered $\sigma_A'\in[0,\lambda[$ and a
further reduced $\sigma_{\rm B}''\in[0,\lambda[$. If $m_{\rm B}'=\intpart(\sigma'_{\rm B}/\lambda)$ is odd, 
a subsequent transformation with 
even $m_{\rm A}'=0$ and even $m_B'+1$ yields an unaltered $\sigma_A'\in[0,\lambda[$ and a reduced 
$\sigma_{\rm B}''\in[\lambda,2\lambda[$.

To conclude, the behavior of quantities in the $\sigma_{\rm A}$-$\sigma_{\rm B}$ plane can be predicted from the knowledge in the
basis unit cell 
\begin{equation}
\mathcal{C}=
\{(\sigma_{\rm A},\sigma_{\rm B})\, |\, 0\le\sigma_{\rm A}<\lambda, \sigma_{\rm A}\le\sigma_{\rm B}<2\lambda\}\,.
\label{eq:unitcell}
\end{equation}
More specifically, knowing the behavior of an observable in this basis unit cell of particle diameters
for all mixing ratios $x$ and particle densities $\rho$, 
one can infer the behavior of
any observable at an arbitrary point in the $\sigma_{\rm A}\sigma_{\rm B}$-plane
at some mixing ratio and density. Figure~\ref{fig:illustration-mapping} illustrates the corresponding mapping.
In the next section, we will demonstrate what this mapping implies for the particle currents, which constitute the primary focus in our study.

Turning back to the single file condition \eqref{eq:single-file-condition}, particle diameters in the basis unit cell violating this condition
are mapped to diameters in another unit cell in the $\sigma_{\rm A}\sigma_{\rm B}$-plane, where the conditions are satisfied.
For example, by carrying out calculations for $\sigma_A'=0.2\lambda$ and $\sigma_B'=1.6\lambda$,
we can predict  the behavior of quantities in
a related system with particle diameters $\sigma_{\rm A}=2.2\lambda$ and  $\sigma_B=3.6\lambda$ satisfying 
$\sigma_{\rm B}<2\sigma_{\rm A}$.

\subsection{Mapping onto system of point particles}
For particle diameters where $\sigma_{\rm A}$ and $\sigma_{\rm B}$ are both even or both odd multiples
of $\lambda$, the mapping to the basic unit cell $\mathcal{C}$ yields $\sigma'_{\rm A}=\sigma'_{\rm B}=0$,
i.e.\ we obtain a system of point particles.

In a system of point particles, collective observables equal those in a system
of independent (noninteracting) particles. Such observables are sums of equivalent terms over all particles,
as it is the case for the total particle density and particle current. For example, for 
$\sigma_{\rm A}=\lambda$, $\sigma_{\rm B}=3\lambda$, $\rho=N/L$,
and any sequence
with $N_\alpha$ particles of type $\alpha$,
the current and the density profile is the same as in the system with $N$ point particles ($\sigma'_{\rm A}=\sigma'_{\rm B}=0$) at density
$\rho'=N/L'=N/[L-N_{\rm A}\lambda-3N_{\rm B}\lambda]$. Hence, the quantities are also the same as in the system of noninteracting particles at density $\rho'$.

Furthermore, the mapping to noninteracting particles allows us to gain a deeper understanding of the
shape of the basic unit cell $\mathcal{C}$ in Eq.~\eqref{eq:unitcell} (see also Fig.~\ref{fig:illustration-mapping}).
In particular, one may wonder why $\sigma_{\rm B}$ in the unit cell is not bounded
by $\lambda$ but by $2\lambda$. 

For hard spheres in one dimension, a
mean interaction force between two particles in contact can be nonzero only
if the mean external forces exerted
on particle clusters left and right of the contact point are different, see \cite{Antonov/etal:2022c} and
the supplemental material of 
Ref.~\cite{Antonov/etal:2022a}. 
If $\sigma_{\rm A}$ and $(\sigma_{\rm A}+\sigma_{\rm B})/2$
are multiple integers of $\lambda$, the external forces on all particles in contact are the same 
(for any position of the particles)
and, accordingly, also the mean external forces exerted
on clusters of neighboring particles in contact. 
This implies that the mean interaction forces vanish. The particles, however, keep their ordering. For obtaining the upper bound for $\sigma_{\rm B}>\sigma_{\rm A}$ in the unit cell, one can
determine the smallest $\sigma_{\rm B}$ leading to zero mean interaction forces if $\sigma_{\rm A}=0$.
This is given by $(\sigma_{\rm A}+\sigma_{\rm B})/2=\sigma_{\rm B}/2=\lambda$, i.e.\ $\sigma_{\rm B}=2\lambda$.

The argument can be generalized to three and more types of hard spheres with diameters 
$\sigma_1<\sigma_2<\ldots<\sigma_q$, $q=2,3\ldots$. Vanishing mean interaction forces are obtained if
$(\sigma_\alpha+\sigma_\beta)/2$ are integer multiples of $\lambda$ for all $\alpha,\beta=1,\ldots,q$. This let us conjecture that the basic unit cell in such a system is 
$\mathcal{C}_q=\{(\sigma_1,\ldots,\sigma_q)\,|\,0\le\sigma_1<\lambda,
\sigma_{\rm 1}\le\sigma_2< 2\lambda,\ldots,\sigma_{q-1}\le\sigma_q<2(q-1)\lambda\}$.

\section{Mapping between currents}
\label{sec:mapping-of-currents}
Due to the single-file condition, both types of particles have the same mean velocity, i.e.\
\begin{equation}
v_{\rm A}=v_{\rm B}=v(x,\rho,\sigma_{\rm A},\sigma_{\rm B})\,.
\end{equation}
The equality~\eqref{eq:path-probabilites} of path probabilities in systems $\mathcal{S}$ and $\mathcal{S}'$
implies that $v(x,\rho,\sigma_{\rm A},\sigma_{\rm B})$
has to obey the relation
\begin{equation}
v(x,\rho,\sigma_{\rm A},\sigma_{\rm B})=v(x,\rho',\sigma_{\rm A}',\sigma_{\rm B}')\,.
\end{equation}
This mapping applies for equal $m_\alpha$ and for different $m_\alpha$ if both $m_\alpha$ are even.

The particle currents $J_\alpha=\rho_\alpha v_\alpha=\rho_\alpha v$ satisfy (note that $\rho_\alpha/\rho_\alpha'=\rho/\rho'$)
\begin{align}
&J_\alpha(x,\rho,\sigma_{\rm A},\sigma_{\rm B})
=\frac{\rho}{\rho'}\,J_\alpha(x,\rho',\sigma_{\rm A}',\sigma_{\rm B}')
\label{eq:mapping-J-1}\\
&\hspace{4em}=\frac{1}{1\!+\![m_{\rm A}(1\!-\!x)\!+\!m_{\rm B}x]\rho'\lambda}J_\alpha(x,\rho',\sigma_{\rm A}',\sigma_{\rm B}')\,.
\nonumber
\end{align}
The total particle current $J=J_{\rm A}+J_{\rm B}=\rho v$ obeys the same transformation rule. 
We can rewrite it as
\begin{equation}
\frac{J(x,\rho,\sigma_{\rm A},\sigma_{\rm B})}{1\!-\![m_{\rm A}(1\!-\!x)\!+\!m_{\rm B}x]\rho\lambda}=J(x,\rho',\sigma_{\rm A}',\sigma_{\rm B}')\,.
\label{eq:mapping-J-2}
\end{equation}
This means that scaled currents for density $\rho$ and different $\sigma_\alpha=\sigma_\alpha'+m_\alpha\lambda$  on the left-hand side of this equation must equal
the current $J(x,\rho',\sigma_{\rm A}',\sigma_{\rm B}')$ for density $\rho'=\rho/(1\!-\![m_{\rm A}(1\!-\!x)\!+\!m_{\rm B}x]\rho\lambda)$ in the basic unit cell $\mathcal{C}$.

Figure~\ref{fig:current-maping} demonstrates the mapping between simulated currents for different $\sigma_\alpha=\sigma_\alpha'+m_\alpha$
where the particle diameters in $\mathcal{C}$
were chosen as $\sigma'_{\rm A}=0.4$ and $\sigma'_{\rm B}=0.7$ ($\lambda=1$ according to our
units, see Sec.~\ref{sec:model}).
The simulations were carried out for a fixed total number $N=40$ of particles and a system length $L'=50$, i.e.\ a density $\rho'=0.8$ if the particle diameters 
are in $\mathcal{C}$ ($m_{\rm A}=m_{\rm B}=0$). The mapping applies to each sequence of the A and B particles. In the figure, it is shown after averaging 
the currents over 40 random sequences for each mixing ratio $x$. If $m_\alpha\ne0$, the densities $\rho$ corresponding to $\rho'$ in Eq.~\eqref{eq:rhoprime-rho} are 
adjusted by changing the system length for fixed $N=40$. 
In Fig.~\ref{fig:current-maping}(a), the particle diameters are increased by equal 
integer multiples of $\lambda$ ($m_{\rm A}=m_{\rm B}=1,2,3$). The currents then fall onto a common master curve given by 
$J(x,\rho'=0.8,\sigma_{\rm A}'=0.4,\sigma_{\rm B}'=0.7)$. 
In Fig.~\ref{fig:current-maping}(b), only the particle diameter $m_{\rm B}$ is increased by integer 
multiples of $\lambda$. In this case, the currents fall onto the master curve $J(x,\rho'=0.8,\sigma_{\rm A}'=0.4,\sigma_{\rm B}'=0.7)$ for even $m_{\rm B}$, and onto the different master curve $J(x,\rho'=0.8,\sigma_{\rm A}'=0.4,\sigma_{\rm B}'=1.7)$ for odd $m_{\rm B}$.

\begin{figure*}[t!]	
\centering
\includegraphics[width=\textwidth]{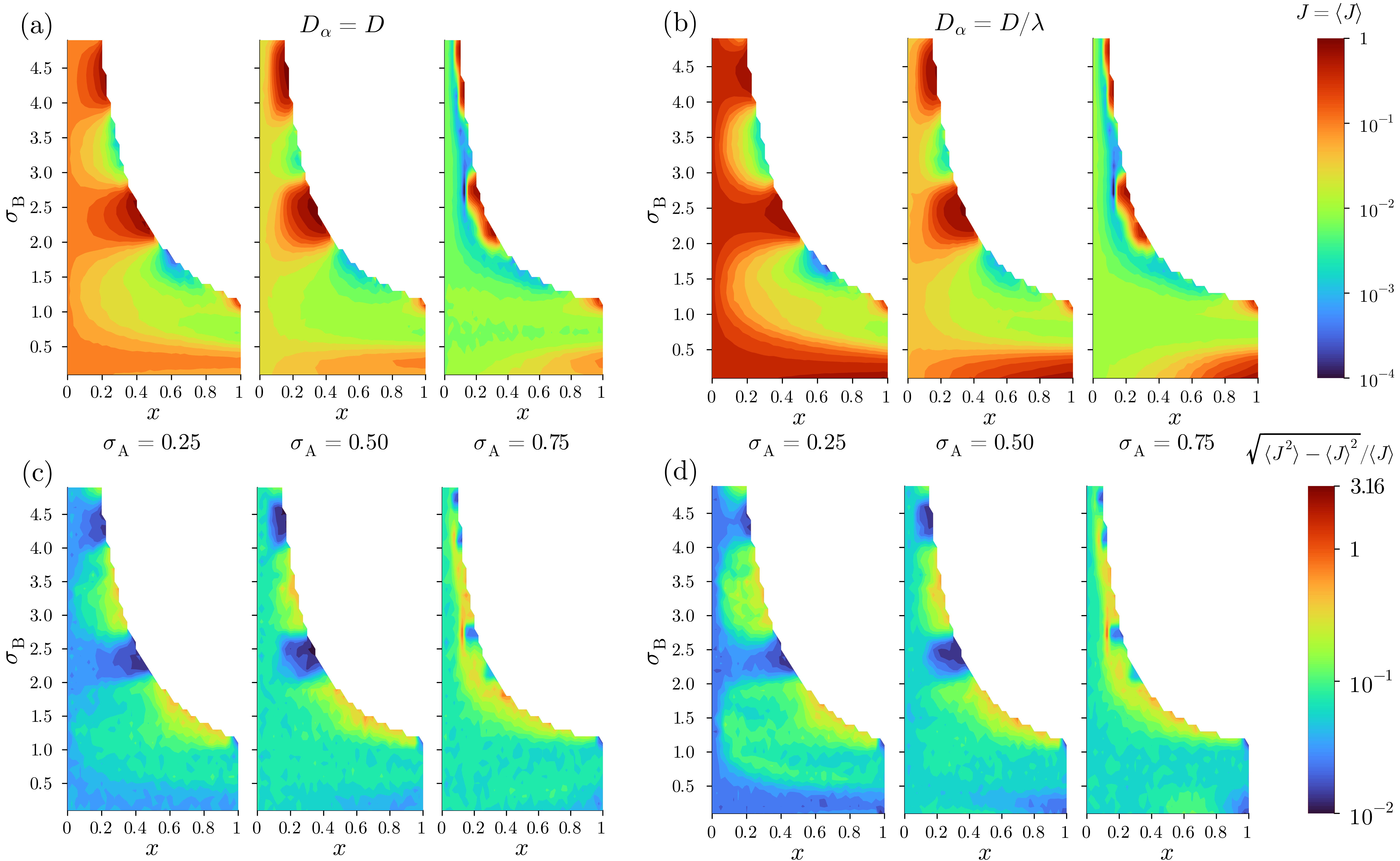}
\caption{Stationary currents [panels (a), (b)] and their relative standard deviation for different orderings of particles [panels (c), (d)] in dependence of the mixing ratio $x$ and particle diameter $\sigma_{\rm B}$ for the parameters sets I 
[$D_\alpha=D$, panels (a),(c)] and
II [$D_\alpha=D/\sigma_\alpha$, panels (b), (d)]  ($\rho=0.8$, three representative values of $\sigma_{\rm A}$).
The magnitudes of the currents and relative standard deviations are given by the colorbars. For completeness, we show results also for $\sigma_{\rm B}<\sigma_{\rm A}$ (which are redundant in principle, as they mirror results for  $\sigma_{\rm A}<\sigma_{\rm B}$).}
\label{fig:currents_and_seq-seq_fluctuations}
\end{figure*}	

\section{Dependence of currents on\newline mixing ratio}
\label{sec:simulated-currents}
In our investigation of stationary particle currents we focus on random mixtures of the two types of particles, i.e.\ random orderings of A and B particles.  The analysis is carried out for various mixing ratios $x$ and particle
diameters $\sigma_{\rm B}$ at a fixed density $\rho=0.8$ and
three fixed particle diameters $\sigma_{\rm A}=0.25$, 0.5, and 0.75. 
We will show results for equal $D_\alpha=D$ and for different $D_\alpha=D/\sigma_\alpha$ according to Eq.~\eqref{eq:Dratio}
(and mobilities $\mu_\alpha=D_\alpha/k_{\rm B}T$). In the following we will refer to the sets of parameters as set I for 
$D_\alpha=D$ and set II for $D_\alpha=D/\sigma_\alpha$. For $x=0$ and $x=1$, the systems become monodisperse ones
for which the behavior of particle currents was discussed in~\cite{Lips/etal:2018, Lips/etal:2019}.

The three $\sigma_{\rm A}$ values are representative for small, intermediate and large diameters of the
smaller A spheres in the basic unit cell. 
The high value of $\rho=0.8$ is chosen to demonstrate the impact
of the hard-sphere interactions on the behavior of particle currents upon mixing of the two particle types.
We have also performed simulations for a small density $\rho=0.2$ and found an analogous, but less pronounced variation of particle currents with $x$ and the particle diameters. This finding let us conclude
that the density is important for the magnitude of current variation with the mixing ratio but not for the overall functional dependence on $x$ and the $\sigma_\alpha$. 

Due to the single-file constraint, the currents are not the same for all particle orderings, and the question arises, 
how strongly currents fluctuate between orderings. To answer this question, we have calculated the
relative standard deviation $\sqrt{\langle J^2\rangle-\langle J\rangle^2}/\langle J\rangle$ ($\langle J\rangle=J$)
as a function of $x$ and $\sigma_{\rm B}$
for parameters sets I and II, using 40 random sequences 
of $N=40$ particles. The results in Figs.~\ref{fig:currents_and_seq-seq_fluctuations}(c) and (d) show 
that the relative standard deviation is not exceeding 10\% for most values of $x$ and $\sigma_{\rm B}$. Where it does,
the current $J=\langle J\rangle$ is very low, as can be seen from Figs.~\ref{fig:currents_and_seq-seq_fluctuations}(a) and (b), 
where we plotted
$J$ as a function of $x$ and $\sigma_{\rm B}$ for parameter sets I and II. If $J$ is very low, even 
small changes of currents with the particle orderings can lead to large relative fluctuations. 

To summarize, except for very low $J$, we can speak about a typical current for a given mixing ratio independent of the specific particle ordering. These typical currents, which we obtain by averaging over random particle orderings at fixed $x$ will be discussed in more detail now.

Comparing the results for parameter sets I [$D_\alpha=D$, Fig.~\ref{fig:currents_and_seq-seq_fluctuations}(a)]
and II [$D_\alpha=D/\sigma_\alpha$, Fig.~\ref{fig:currents_and_seq-seq_fluctuations}(b)],
the location and overall pattern
of regions with high and low currents in the $x\sigma_{\rm B}$-plane are similar. 
The main effect induced by the varying diffusion
coefficients is to weaken differences in the current magnitudes.
This demonstrates the minor impact of varying diffusion coefficients on current changes.

The mixing of A and B particles can lead to strong changes of the current by many orders of magnitude,
note the logarithmic scale bars in
Figs.~\ref{fig:currents_and_seq-seq_fluctuations}(a) and (b). 
Our identification of a basic unit cell 
$\mathcal{C}$ in
Sec.~\ref{sec:mapping-of-currents} is helpful to gain a first understanding of the patterning of regions with high and low currents.
It implies a repetitive behavior of currents when $\sigma_{\rm B}$ is changed by integer multiples of $2\lambda$,
and this is indeed reflected in Fig.~\ref{fig:currents_and_seq-seq_fluctuations}. Namely, regions of maxima (dark orange) 
appear at $\sigma_{\rm B}$ close to
$2.4$  and $4.4$, and regions of minima of the current (green-blue) appear at $\sigma_{\rm B}$ close to
$1.7$  and $3.7$.
Although there are differences in detail, the minima and maxima occur at similar positions for the three distinct
$\sigma_{\rm A}$ values in the basic unit cell.

The minima and maxima are most pronounced close to
the full coverage, which is given by the maximal possible $x=x_{\rm max}$ in the inequality \eqref{eq:x-constraint}, i.e.\
for $\sigma_{\rm B}\ge1/\rho$ by
\begin{equation}
x_{\rm max}=\frac{1/\rho-\sigma_{\rm A}}{\sigma_{\rm B}-\sigma_{\rm A}}\,,
\end{equation}
where $1/\rho=1.25$ for $\rho=0.8$ in Fig.~\ref{fig:currents_and_seq-seq_fluctuations}. 
When decreasing $x$ at fixed 
$\sigma_{\rm B}$ in Fig.~\ref{fig:currents_and_seq-seq_fluctuations}, the currents in general increase (decrease) if
they have a minimum (maximum) close to $x_{\rm max}$.
Hence, for understanding the overall variation of currents with $x$ and $\sigma_{\rm B}$ in the colormaps 
in Figs.~\ref{fig:currents_and_seq-seq_fluctuations}(a) and (b), we can focus
on explaining the occurrence of the minima and maxima appearing near $x_{\rm max}$.

\begin{figure}[b!]
\centering
\includegraphics[width=\columnwidth]{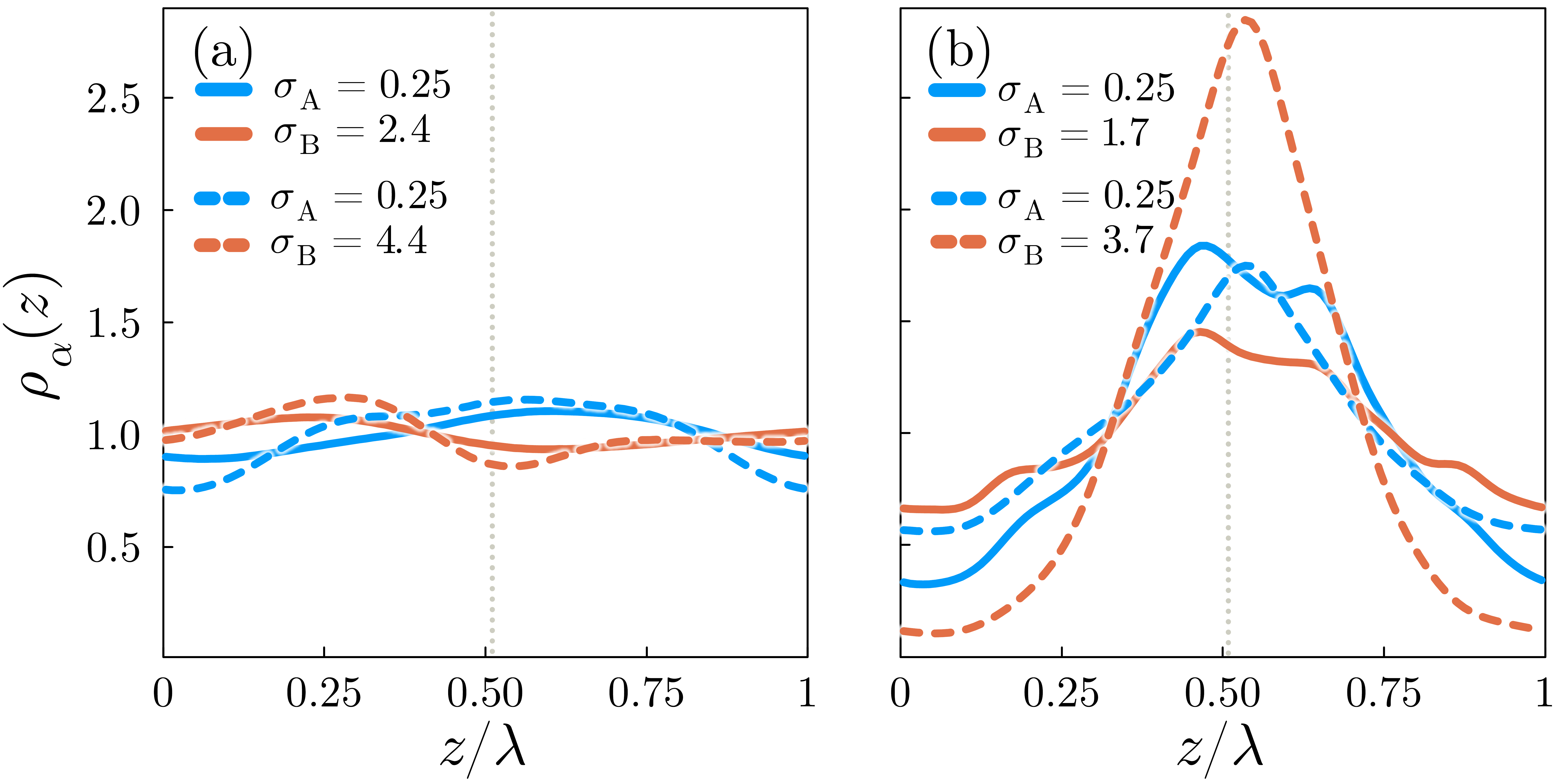}
\caption{Averaged density profiles for tight particle packings. In (a) particles displace each other greatly, while in (b), the particles are confined in to potential wells blocking each other from motion. 
Gray dashed line indicates potential minimum. }
\label{fig:densities}
\end{figure}

Let us first look at the local densities 
$\rho_\alpha(z)=\langle\sum_i\delta(z-z_i(t))\rangle/N_\alpha$ of particles of type $\alpha$,
where the sum runs over all particles of type $\alpha$. We determined $\rho_\alpha(z)$
from the simulations by calculating the mean number of $\alpha$ particles falling into a small bin around the position $z$. 

Figure~\ref{fig:densities} shows $\rho_\alpha(z)$
for parameter set I ($D_\alpha=D$), $\sigma_{\rm A}=0.25$, for (a)
$\sigma_{\rm B}=2.4$ and 4.4 [i.e., maxima in Fig.~\ref{fig:currents_and_seq-seq_fluctuations}(a)], and (b)
$\sigma_{\rm B}=1.7$ and 3.7 [minima in Fig.~\ref{fig:currents_and_seq-seq_fluctuations}(a)].
In Fig.~\ref{fig:densities}(a),
the profiles are nearly flat, while in Fig.~\ref{fig:densities}(b), the profiles show pronounced maxima
close to minima of the external cosine potential $U(z)$ [Eq.~\eqref{eq:external-potential}], 
or, more precisely, minima of the
locally tilted potential $U(z)-fz$, which is indicated by the dashed vertical line in Fig.~\ref{fig:densities}.
The flat shape indicates an enhancement of barrier crossing rates for individual particles by an effective barrier 
reduction: the particles are strongly displaced from the minima of the external potential and 
need to surmount lower barriers. 

\begin{figure}[t!]
\centering
\includegraphics[width=0.8\columnwidth]{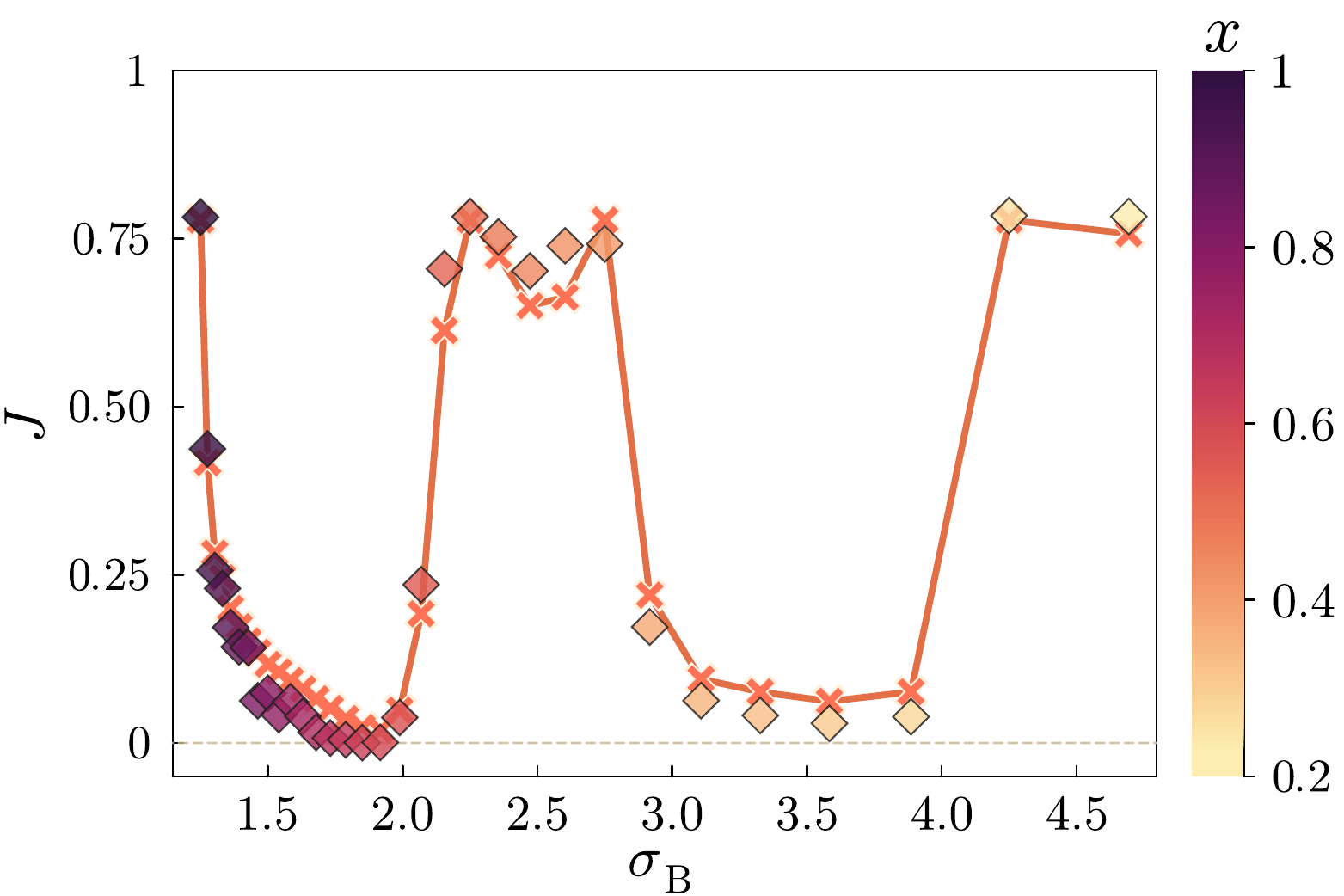}
\caption{Mean particle current $J(x_{\rm max},\rho,\sigma_{\rm A},\sigma_{\rm B})$ at full coverage as a function of
$\sigma_{\rm B}$ for fixed $\rho=0.8$ and $\sigma_{\rm A}=0.25$ (solid line with crosses) calculated from  Eq.~\eqref{eq:Jfullcoverage}. An average over $5\times10^4$ different particle ordering has been carried out.
The diamonds indicate simulations results for the same parameters $\rho=0.8$ and $\sigma_{\rm A}=0.25$ at almost full
coverage (see text).}
\label{fig:comparison_Jcalc-Jsim}
\end{figure}

Can we explain why the current maxima close to $x_{\rm max}$ occur at $\sigma_{\rm B}\simeq2.4$ and 4.4, while
the minima occur at $\sigma_{\rm B}\simeq1.7$ and 3.7? This can be done by considering that in the limit of full coverage, all particles in the system are in contact, i.e.\ they form a single particle cluster, and 
the current can be determined by the motion of this single cluster. 
For any given ordering of the particles, the equation of motion for the center of mass position $z_{\rm cm}$ of the cluster is
\begin{equation}
\frac{\dd z_{\rm cm}}{\dd t} =\mu f-\frac{\mu}{N} \sum_{i=1}^N U'(z_i) + \frac{\sqrt{2D}}{N}\,\sum_{i=1}^N\xi_i(t)\,,
\label{eq:langevin-zcm}
\end{equation}
where
\begin{equation}
z_{\rm cm}=\frac{1}{N}\sum_{i=1}^Nz_i=
z_1+\frac{1}{N}\sum_{i=2}^N\sum_{j=1}^{i-1}\sigma_{j,j+1}
\end{equation}
and $\sigma_{j,j+1}=(\sigma_j+\sigma_{j+1})/2$. Accordingly,
we can use $z=z_1$ for defining the cluster position. Its time evolution is given by
the Langevin equation
\begin{equation}
\frac{\dd z}{\dd t} =\tilde\mu\, \left(Nf-\frac{\dd u_{\rm cm}}{\dd z}\right)+\sqrt{2\tilde D}\,\xi(t)\,,
\label{eq:langevin-z}
\end{equation}
where $\tilde\mu=\mu/N$, $\tilde D=D/N=k_{\rm B}T\tilde\mu$ and $\xi(t)$ is a Gaussian noise with $\langle\xi(t)\rangle=0$ and
$\langle\xi(t)\xi(t')\rangle=\delta(t-t')$. For a particle ordering with diameters $\sigma_1,\ldots,\sigma_N$
the potential for the center of mass motion in Eq.~\eqref{eq:langevin-z} is
\begin{align}
u_{\rm cm}(z)&=u_{\rm eff}(z;\sigma_1,\ldots,\sigma_N)=\sum_{i=1}^NU\Bigl(z+\sum_{j=1}^{i-1}\sigma_{j,j+1}\Bigr)
\nonumber\\
&=\frac{|A|U_0}{2}\cos(\frac{2\pi z}{\lambda}+\varphi)\,,
\label{eq:ueff}
\end{align}
where $A\!=\!\sum\limits_{j=1}^N \exp(2\pi i\sum\limits_{k=1}^{j-1}\sigma_{k,k+1}/\lambda)$ and $\varphi=\arg(A)$. This 
potential is $\lambda$-periodic. The mean velocity $\bar v$
of the Brownian motion of a single particle in a periodic potential under a constant drag force
is known exactly \cite{Stratonovich:1958, Ambegaokar/Halperin:1969, Reimann:2002}. For
the stationary current $\rho\bar v$ we obtain
($\beta=1/k_{\rm B}T$) 
\begin{align}
&j(x_{\rm max},\rho;\sigma_1,\ldots,\sigma_N)=\label{eq:Jfullcoverage}\\
&\hspace{0.5em}\frac{D}{N}\,\rho\,\frac{\lambda(1-e^{-\beta Nf \lambda})}
{\int\limits_0^\lambda \dd z\! \int\limits_z^{z+\lambda}\! \dd y\, 
\exp[\beta(u_{\rm cm}(y)\!-\!Nfy\!-\!u_{\rm cm}(z)\!+\!Nfz)]}\,.\nonumber
\end{align}
By averaging this current over particle orderings, we obtain 
$J(x_{\rm max},\rho,\sigma_{\rm A},\sigma_{\rm B})=\langle j(x_{\rm max},\rho,\sigma_1,\ldots,\sigma_N)\rangle$, 
$\langle\ldots\rangle$ denoting the average over orderings. As an example, we show in Fig.~\ref{fig:comparison_Jcalc-Jsim} 
the so-calculated current at full coverage as a function of $\sigma_{\rm B}$ for fixed
$\rho=0.8$ and $\sigma_{\rm A}=0.25$ (solid line with crosses). In this numerical calculation, we have taken an average of $5\times10^4$ 
different particle orderings.
The calculated values agree very well with simulation results that we obtained by averaging over 100~orderings.
To check that the analytical results for full coverage are describing the limiting behavior of the many-particle system, we performed these 
simulations by taking the parameters at full coverage and then adding one potential well to give the particles some free space to move.
The corresponding simulation results indicated by the diamonds in Fig.~\ref{fig:comparison_Jcalc-Jsim} are well 
described by the analytical prediction for full coverage.

Further insights into the behavior of $J(x_{\rm max},\rho,\sigma_{\rm A},\sigma_{\rm B})$ can be gained by 
averaging the amplitude factor $|A|$, yielding the 
effective potential 
\begin{equation}
\label{eq:UN}
U_{\rm eff}(z)=\frac{\langle |A|\rangle U_0}{2}\cos(\frac{2\pi z}{\lambda})\,.
\end{equation}
Note that we did not average $u_{\rm cm}(z)$, as the different phases $\varphi$ for the orderings would
lead to interference effects. These phases, however, are irrelevant for the stationary currents.

\begin{figure}[t!]
\centering
\includegraphics[width=\columnwidth]{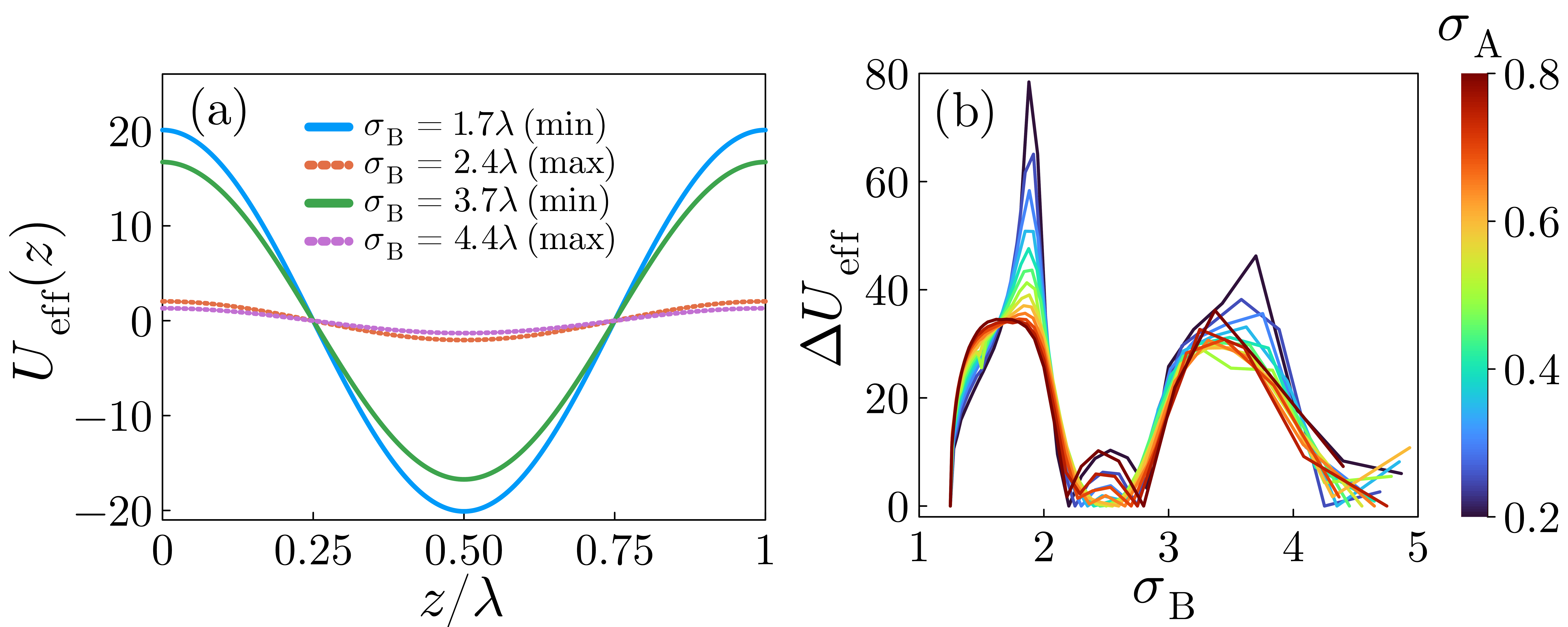}
\caption{(a) Effective potential $U_{\rm eff}(z)$ for the single 
cluster formed by all $N$ particles in the limit of full coverage at values of $\sigma_{\rm B}$
corresponding to minima and maxima of the current in Figs.~\ref{fig:currents_and_seq-seq_fluctuations}(a) and (b). (b) Heights of the potential barrier to be surmounted by the cluster as a function of the particle 
diameter $\sigma_{\rm B}$ for various $\sigma_{\rm A}$.}
\label{fig:effective-potential}
\end{figure}

Figure~\ref{fig:effective-potential}(a) shows $U_{\rm eff}(z)$ for the diameters $\sigma_{\rm B}$ corresponding to
minima and maxima of the current. For $\sigma_{\rm B}=1.7$ and 3.7,
the cluster has to surmount high potential barriers giving rise to
low currents. By contrast, for $\sigma_{\rm B}=2.4$ and 4.4 no potential barriers can be seen on the scale of the
$U_{\rm eff}$ axis in the figure, i.e.\ potential barriers are negligible compared to $k_{\rm B}T$,
leading to high currents.

Closer inspection of $U_{\rm eff}(z)$ allows us to identify the particle diameters leading
to highest and lowest barrier heights at full coverage. For the density $\rho=0.8$, these
barrier heights 
\begin{equation}
\Delta U_{\rm eff}=\langle |A|\rangle U_0, 
\end{equation}
are shown in Fig.~\ref{fig:effective-potential}(b)
as a function of $\sigma_{\rm B}$ for various $\sigma_{\rm A}$. Low and high barriers occur at similar
$\sigma_{\rm B}$ for different $\sigma_{\rm A}$, which explains the aforementioned insensitivity of 
the locations of regions with low and high current in the $\sigma_{\rm A}\sigma_{\rm B}$-plane [see the discussion of
Figs.~\ref{fig:currents_and_seq-seq_fluctuations}(a) and (b)].

\section{Summary and Conclusion}
We have studied Brownian transport of a binary mixture of hard-spheres A and B with two different diameters
$\sigma_{\rm A}$ and $\sigma_{\rm B}\ge\sigma_{\rm A}$. The particles are driven
through a periodic potential with wavelength $\lambda$ under single-file conditions, where they cannot overtake each other. By extensive Brownian 
dynamics simulations we calculated particle currents for a wide range of mixing ratios $x$ and particle diameters at several particle densities. 
A complex behavior in dependence of the mixing ratio and particle diameters was obtained, with currents varying by many orders of magnitude at high particle densities. 
Fluctuations of the currents between
different particle reorderings turned out to be small compared to the mean values, except for
the smallest particle currents.
The strong variations of currents are only weakly affected when taking into account a scaling of mobilities with the particle diameter according to Stokes' friction law. 

To understand the complex behavior, one can thus consider both particle types to have the same mobility. For that situation, we derived an exact 
mapping between probabilities of Brownian paths of different systems, where a
system is related to another one 
with hard-sphere diameters and system lengths differing by integer multiples 
of the potential wavelength. As a consequence, there exists a basic unit cell in the
$\sigma_{\rm A}\sigma_{\rm B}$-plane, with $\sigma_{\rm A}$ lying between zero and $\lambda$, and
$\sigma_{\rm B}$ between $\sigma_{\rm A}$ and $2\lambda$.
Once the behavior of an observable is determined within the basic unit cell for all particles at a specific mixing ratio, it can be inferred for any combination of particle diameters and density at the same mixing ratio.

Application of the mapping to particle currents allowed us to understand that minima and maxima of them occur repeatedly in the $\sigma_{\rm A}\sigma_{\rm B}$-plane at $\sigma_{\rm B}$ values separated
by approximately two wavelengths. We further showed that the overall current variation
 with $x$ and with
the hard-sphere diameters $\sigma_{\rm A}$ and $\sigma_{\rm B}$ can be inferred essentially from the current change with particle diameters in the limit of full coverage, when
the sum of all particle diameters equals the system size. This change of current in the limit of full coverage
can be explained from a calculation of an effective potential for particle transport. 

The methods developed here for understanding Brownian dynamics of a binary hard-sphere mixture (mapping between Brownian paths, effective potential at full coverage) can be applied to arbitrary periodic 
potentials and extended to mixtures of more than two types of hard spheres. Following earlier studies for 
monodispersive systems \cite{Ryabov/etal:2019}, it will be interesting to explore how
the collective transport properties are reflected in local transition dynamics of tagged particles.
A further interesting extension is the investigation of systems, where the number of particles exceeds the number of potential wells. In such overcrowded systems, we expect that the particle transport is dominated by particle cluster waves, which at low temperatures appear as stable propagating solitons \cite{Antonov/etal:2022a}. 

More elaborate analytical treatments may be possible in the future, as, for example,
by approximate treatments of exact equations for the time evolution of particle densities and correlations derived from the many-body Smoluchowsky equation \cite{Lips/etal:2018}, by using dynamical density functional theory for hard-sphere mixtures \cite{Stopper/etal:2015, Stopper/etal:2018}, or power density functional theory \cite{Schmidt/Brader:2013, Schmidt:2022}. Based on earlier investigations \cite{Lips/etal:2020, Antonov/etal:2021}, we believe that an understanding of hard-sphere mixture dynamics in periodic structures will provide a suitable basis for treating systems with other particle interactions. 

Our theoretical treatment of driven Brownian motion of a binary mixture 
could lay a suitable basis also to understand anomalous uphill diffusion against a concentration gradient
of one particle type, while the other type of particle behaves normally. This effect was experimentally observed in two-component diffusion through
nanopores \cite{Lauerer/etal:2015} and was related to cross-coupling terms (nondiagonal Onsager coefficients), i.e.\ that a concentration gradient in one component influences the current of the other component.  We believe that nondiagonal Onsager coefficients are strongly affected by changes of the particle size.

\section*{Acknowledgements}
Financial support by the Czech Science Foundation (Project No.\ 20-24748J) and the Deutsche Forschungsgemeinschaft (Project No.\ 432123484) is gratefully acknowledged. Computational resources were provided by the e-INFRA CZ project (ID:90140), supported by the Ministry of Education, Youth and Sports of the Czech Republic. D.V.\ acknowledges the support by the Charles University (project Primus/22/SCI/017). 

\section*{Data availability statement}
The data that support the findings of this study are available from the corresponding author upon reasonable request.

%

\end{document}